\documentclass[journal ]{new-aiaa}
\usepackage[utf8]{inputenc}
\usepackage{textcomp}
\usepackage{graphicx}
\usepackage{subcaption} 
\usepackage{amstext} 
\usepackage{amsfonts}
\usepackage[font={footnotesize}]{caption}
\usepackage{hyperref}
\usepackage{xcolor}
\usepackage{tikz}
\usepackage[utf8]{inputenc}
\hypersetup{colorlinks=true,urlcolor=blue,linkcolor=black,citecolor=black} 
\usepackage{amsmath}
\usepackage[version=4]{mhchem}
\usepackage{siunitx}
\usepackage{longtable,tabularx}
\usepackage{lineno} 
\usepackage{soul} 
\title{The Peregrine Falcon's Dive: On the Pull-Out Maneuver and Flight Control Through Wing-Morphing}

\author{Omar Selim \footnote{PhD Student, Department of Mechanical Engineering and Aeronautics, omar.selim@city.ac.uk}} 
\affil{City University, of London, London, EC1V 0HB, UK}
\author{Erwin. R. Gowree\footnote{Associate Professor, Department of Aerodynamics, Energetics and Propulsion, erwin-ricky.gowree@isae-supaero.fr}}
\affil{ISAE-SUPAERO, Université de Toulouse, 31055, France}
\author{Christian Lagemann\footnote{PhD Student, Institute of Aerodynamics, c.lagemann@aia.rwth-aachen.de}}
\affil{RWTH Aachen University, Aachen, 52062, Germany}
\author{Edward Talboys \footnote{PhD Student, Department of Mechanical Engineering and Aeronautics, edward.talboys.1@city.ac.uk}, Chetan Jagadeesh \footnote{Lecturer, Department of Mechanical Engineering and Aeronautics, chetan.jagadeesh.1@city.ac.uk} and Christoph Br\"{u}cker \footnote{Professor, Department of Mechanical Engineering and Aeronautics, christoph.bruecker@city.ac.uk}}
\affil{City University, of London, London, EC1V 0HB, UK}

\begin{document}

\maketitle
\small\begin{abstract}
\textbf{
During the pull-out maneuver, Peregrine falcons were observed to adopt specific flight configurations which are thought to offer an aerodynamic advantage over aerial prey. 
Analysis of the flight trajectory of a falcon in a controlled environment shows it experiencing load factors up to 3 and further predictions suggest this could be increased up to almost 10g during high-speed pull-out.
This can be attributed to the high maneuverability promoted by lift-generating vortical structures over the wing.
Wind-tunnel experiments on life-sized models together with high fidelity simulations on idealized models, which are based on taxidermy falcons in different configurations, show that deploying the hand-wing in a pull-out creates extra vortex-lift, similar to that of combat aircraft with delta wings. 
The aerodynamic forces and the position of aerodynamic center were calculated from Large Eddy Simulations of the flow around the model. This allowed for an analysis of the longitudinal static stability in a pull-out, confirming that the falcon is flying unstably in pitch with a positive slope in the pitching moment and a trim angle of attack of about 5$^\circ$, possibly to maximize responsiveness. The hand-wings/primaries were seen to contribute to the augmented stability acting as `elevons' would on a tailless blended-wing-body aircraft.
} 
\end{abstract}
\pagebreak

\section*{Nomenclature}


{\renewcommand\arraystretch{1.0}
\noindent\begin{longtable*}{@{}l @{\quad=\quad} l@{}}
$AC$  & aerodynamic center \\
$AR$  & aspect ratio, defined as $b^2/S$ \\
$\alpha$  & angle of attack \\
$\alpha_t$  & trim angle of attack \\
$b$  & wingspan, [m] \\
$C$  & characteristic chord length, [m] \\
$C_D$  & drag coefficient \\
$C_L$  & lift coefficient \\
$C_{L_p}$ &    potential lift coefficient \\
$C_{L_T}$  & theoretical lift coefficient \\
$C_{L_v}$  & vortex lift coefficient \\
$C_{m_0,ac}$  & zero-lift pitching moment coefficient about aerodynamic center \\
$C_{m,cg}$  & pitching moment coefficient about center of gravity \\
$C_w$  & wing max chord, [m] \\
$C_{(x,y,z)}$  & force coefficient in the \textit{x,y,z} direction \\
$F_{(x,y,z)}$  & force in the \textit{x,y,z} direction, [N] \\
$g$  & acceleration of gravity, $[ms^{-2}]$ \\
$K_p$  & potential flow coefficient \\
$K_v$  & vortex flow coefficient \\
$\Lambda$  & leading edge wing sweep \\
$n$  & load factor \\
$q_\infty$  & free-stream dynamic pressure, [Pa] \\
$S$  & characteristic planform area, [$m^2]$ \\
$\theta$  & angle of pitch [$rad$] \\
$\dot{\theta}$ & rate of pitch, [$rad.s^{-1}$] \\
$x_{cg}$  & position of center of gravity, [m] \\
$W$ & Weight, [N] \\
\multicolumn{2}{@{}l}{Subscripts}\\
cg & center of gravity\\
ac & aerodynamic centre\\
\end{longtable*}}

\pagebreak

\section{Introduction} 
\begin{figure}[t!]
 \centering 
\includegraphics[scale=0.3]{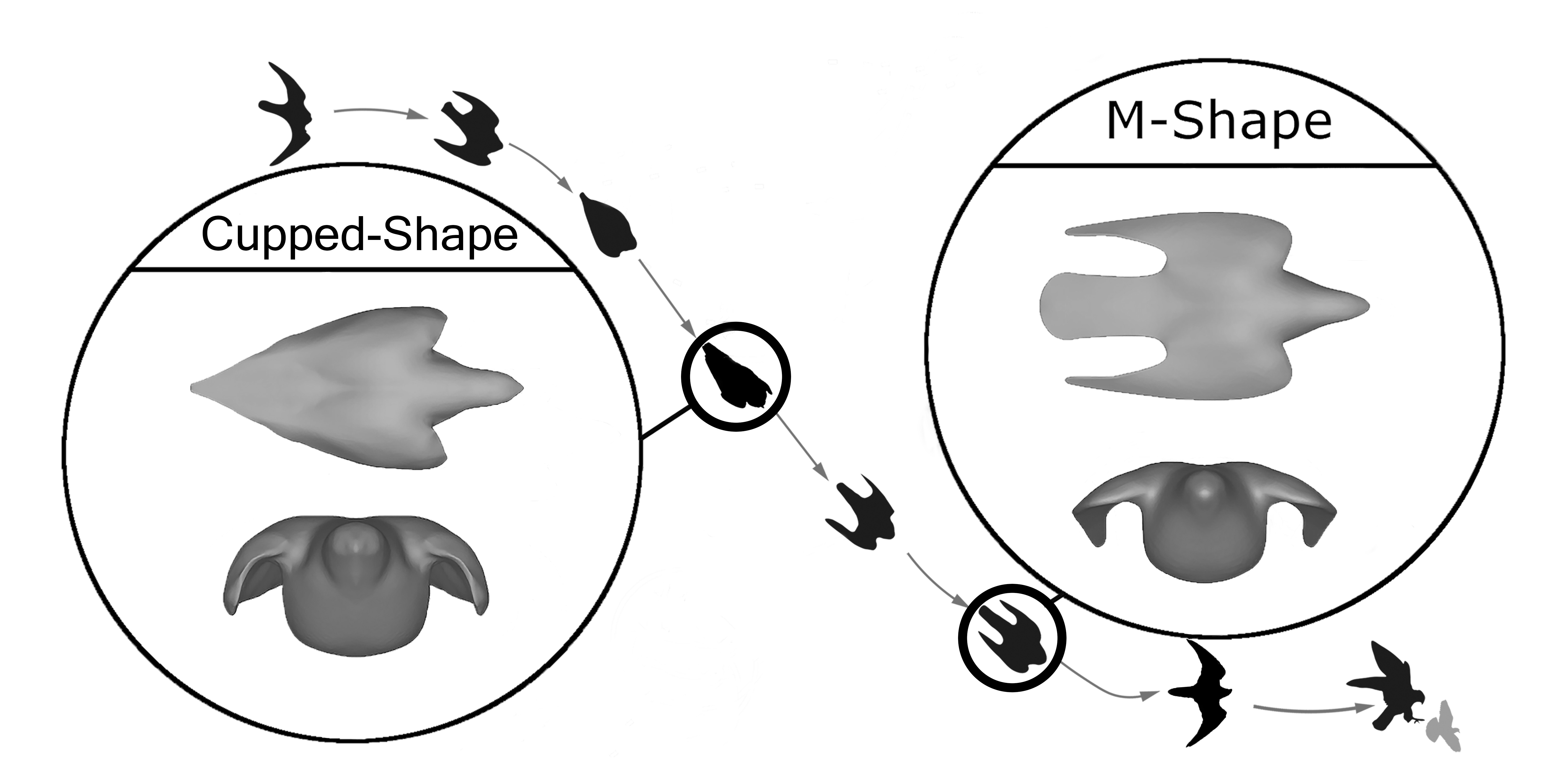}
 \caption{\textbf{The morphological transformation of a Peregrine falcon at various stages during stoop and pull-out. On the side, the elevated and plan views of the corresponding models generated by laser scanning of taxidermied birds and then reproduced in Computer Aided Design.}}
\label{fig:Flight_path}
\end{figure}

\lettrine{T}{he} Peregrine falcon (falco peregrinus) attacks its prey by rapid strike while performing high speed stoops (dives), due to its ability to undergo a variety of morphological transformation. 
It climbs to high altitude during thermal soaring with its large wings completely spread, but the most interesting morphological transformation happens during the stoop, shown in Fig. \ref{fig:Flight_path}.
Although the high speed achieved in the Teardrop-shape \footnote{One of the configurations adopted in a stoop, these different configurations are discussed on page 4} \citep{Alerstam1987, Tucker1998a, Tucker1998b} has so far attracted most of the attention, during the stoop the falcon also shows additional impressive aerobatic performance in other flight configurations.
For instance, while re-adjusting its attitude to increase strike precision, the falcon will open the hand-wings laterally morphing into what is currently designated the Cupped-shape \citep{Alerstam1987}.
This is also employed for slight deceleration.
In this configuration the lateral forces on the wings can be 3 times the weight of the bird and Peregrine falcons can withstand such high loads due to their superior musculo-skeletal structure compared to other avians \citep{schmitz2018}. 
In the final phase immediately before or during prey strike the bird will enter the `pull-out' in which it will morph into what is designated the `M-shape'.
In this configuration, the bird is able to generate considerably more lift without the typically associated increase in drag expected for a classical wing, enabling controlled  flight at moderately high speeds.
These morphological adaptations have been video-captured using sophisticated high-speed recording systems during field experiments \citep{ponitz2014} and wind tunnel analysis has provided us with further evidence of the flow behavior that allows the falcon to achieve such impressive maneuverability. 
This skill is also demonstrated during courtship flights where the falcon can achieve even higher speeds during the dive since it does not have to track a prey \citep{Bleckmann2015}.

\begin{figure}[h!]
 \centering
 \includegraphics[scale=0.6]{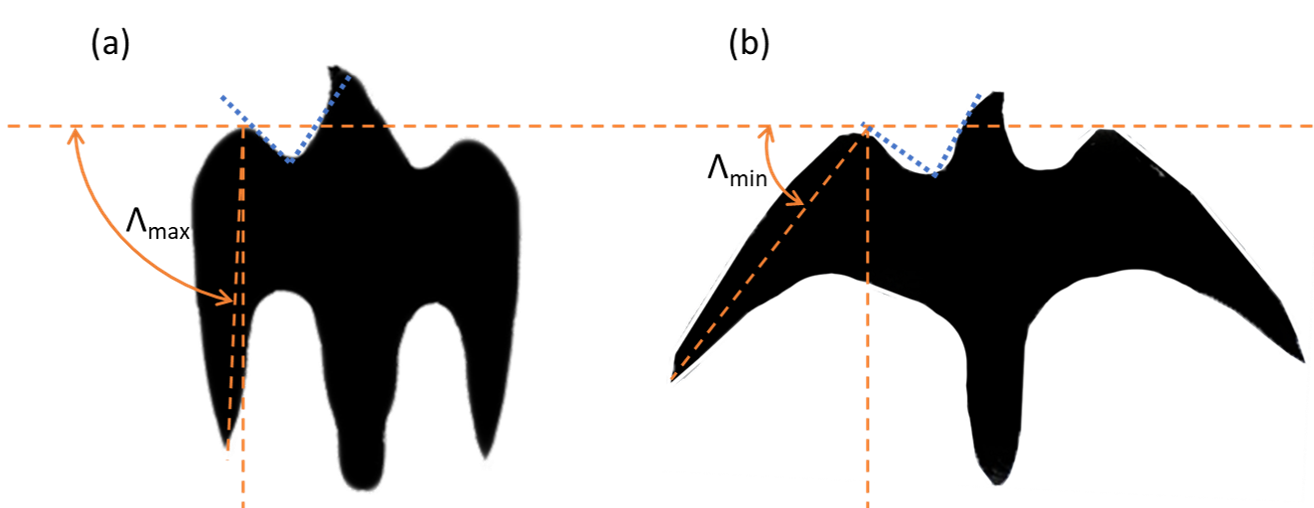}
 \caption{\textbf{Schematic of sweep change (\protect\begin{tikzpicture}
\protect\tikz[baseline=-1pt] \protect\draw[dashed,very thick,color=orange] (0.0,0.0) -- (0.5,0.0);
\protect\end{tikzpicture}) in M-shape configuration illustrating the consistent in-board forward sweep throughout (\protect\begin{tikzpicture}
\protect\tikz[baseline=-1pt] \protect\draw[dashed,very thick,color=blue] (0.0,0.0) -- (0.5,0.0);
\protect\end{tikzpicture})}}
 \label{fig: PlanfCh2}
\end{figure}

\pagebreak
According to \citet{ponitz2014} the falcon’s stoop and early pull-out can be categorized into a succession of four characteristic configurations \citep{Tucker1998b}:
\begin{itemize}
    \item Teardrop-shape
    
This is the configuration typically adopted for maximum speed in the beginning of a stoop.
It is characterized by the wings being completely retracted into the body and the tail remaining completely furled.
The bird has little to no control at this point and is rather similar to an unguided projectile optimizing speed.
    \item Cupped-shape
    
This is the configuration that the bird will adopt in order to adjust its trajectory at high speeds and is accompanied by a marginal deceleration.
It is characterized by the trailing edge of the wings remaining attached to the tail and the leading edge morphing laterally away from the body to a position that creates a `cup' of air between them.
    \item Open-Cupped
    
Somewhat of a bridge between Cupped and M-shape, this configuration is seen to be intermittently adopted during a stoop for control purposes.
It is characterized similar to Cupped-shape but with the leading edge further extended eliminating the `cup' between the wing and body.
    \item M-shape
    
Usually the final configuration adopted in which the bird is `pulling out' of a dive.
In this configuration the bird will keep the in-board section of the wing pushed forward while swinging the out-board section (hand-wing) to {vary the sweep angle and this extension increases the total lift produced.}
It is characterized by the trailing edge of the wing detaching from the body/tail, a forward sweep in the in-board section, and a  reduction in aft-sweep ($\Lambda$) in the out-board section typically ranging from $\Lambda_{min}\approx 40^\circ$ to $\Lambda_{max}\approx 90^\circ$, as shown in figure \ref{fig: PlanfCh2}.
\end{itemize}

\vskip 8mm

Each of these phases was observed not to be simply a fixed position in which the bird will robotically morph but rather a smooth, reversible transition between the phases with certain dominating characteristics exhibited categorising each configuration into which the bird has evolved to morph for performance optimisation, stability augmentation, and/or comfort physiological limits such as maximum shoulder torque \citep{schmitz2018peregrine} \citep{durston2019avian}.

In a stoop and pull-out maneuver, the falcon will be adopting one of the above configurations throughout.
During live flight analysis the bird is seen to adjust its trajectory or correct its attitude very rapidly in Cupped-shape and M-shape in order to increase the chances of striking the prey. 
This is enabled by the excellent roll and yaw control abilities it possesses.
What is peculiar is how the tail remains furled with little to no change in a pull-out in these recordings suggesting that the mechanics of such maneuvers are entirely or almost entirely controlled by the wings in that configuration, similar to what delta wings can achieve when equipped with elevons at the trailing edge. 
As opposed to conventional aircraft the bird does not have a fin and a rudder for lateral control, and therefore uses the wing-tips and the tail to achieve these maneuvers. 
This is confirmed from the live recordings reported in reference \citep{Gowree2018, ponitz2014a, NatGEo1, NatGEo2} where the bird is seen to open-up its wings laterally, sometimes even close to the M-shape during the high speed dive, however it tucks them back in into the Teardrop-shape immediately after to reduce the drag.

Primarily, the falcon’s wing morphing when in M-shape configuration serves to vary the amount of lift produced either to reach a specific load factor – and subsequently a specific radius of pull-out, or to kill lift in order not to exceed the maximum tolerable bending torque about the bird’s shoulder \citep{tucker1970aerodynamics, schmitz2018peregrine}.
By analysing the flight trajectory of a falcon in pull-out, this paper will strive to calculate the limits of these forces throughout the maneuver and estimate the maximum forces the bird would need to withstand with its superior flight speeds.
Secondarily however, the morphing serves as positional adjustments of the center of pressure either spanwise in order to reduce the bending torque about the shoulder girdle without sacrificing lift or longitudinally as pitching moment control augmenting the limited inputs from the furled tail.
This fore and aft movement due to longitudinal perturbations is not unique to falcons and was observed also in hawks and pigeons but often coupled with inputs from a deployed tail \citep{brown1963flight,tucker1992pitching}.

Most fixed wing/forward thrust UAV's/MAV's airborne today fly in a statically stable manner.
This is due to the benefits of the passive stability: namely, that the aircraft does not require input from the controller or from a flight control computer to respond to small longitudinal perturbations.
However, associated with static stability is a limitation of maneuverability.
Naturally, a statically stable platform will offer some passive resistance to departure from equilibrium - an undesirable trait of a nimble predator.
Conversely, modern fighter aircraft will fly in a statically unstable or neutrally stable configuration such as to exploit the benefits in maneuverability.
This kind of flight would not be possible unless controlled by a flight control computer (FCC) able to detect and counter departures from equilibrium instantaneously.

As a continuation of the previous work by \citet{Gowree2018,ponitz2014,ponitz2014a}, the present paper analyses the aerodynamics and mechanics of the flight of a falcon in pull-out maneuver with strong wing-morphing, applying classical flight stability criteria in order to draw parallels with current state-of-the-art highly maneuverable flight demonstrators and to explore the possibility of incorporation of the morphing mechanics and control mechanisms in modern Micro-Air-Vehicles (MAVs) and Unmanned Aerial Vehicles (UAVs).
\pagebreak
\section{Methods}

\noindent\subsection{Life-size Models.} \label{sec:model}
The geometries of the different wing morphing states were already used in our previous studies and were qualified as good representatives of Peregrine falcons in flight \citep{Gowree2018, ponitz2014}. They were derived from a combination of high-resolution digital photographs simultaneous from different perspectives in field-experiments and 3D laser scanning of stuffed animals reconstructed for different flight situations. 
The scanned geometries were imported into a computer-aided design package (CAD) and with the help of the detailed flight images \citep{ponitz2014}, the different wing morphology could be realized with good match to the images from different perspectives. 
Physical models were manufactured using 3D printing techniques for the use in wind-tunnel studies, coated in a matte black paint in order to increase contrast for the flow visualisation. 
The corresponding 3D-printed models are shown in Fig. \ref{fig:Flight_path}. 
The same geometries were exported as STL-type surface meshes to represent the body in a digital form for the LES simulations.

\noindent\subsection{Simulations}
\begin{figure}[h!] 
\centering\includegraphics[scale = 0.85]{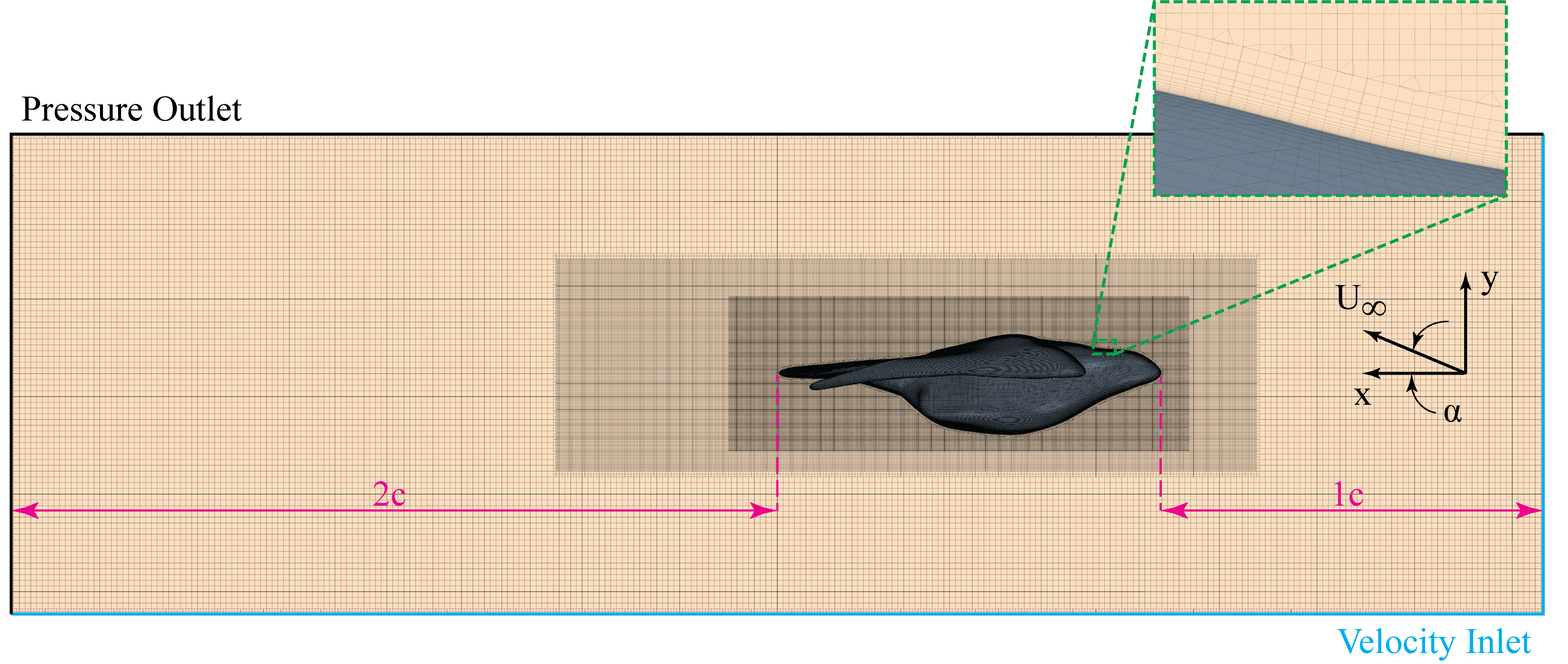}
\caption{\textbf{CFD computational domain viewed in the x-y plane. The boundary conditions, are indicated as Pressure outlet in black and the velocity inlet in blue, the walls of the domain in the z-plane are not indicated on the figure but they are both pressure outlets. A close up view of the near wall mesh can be seen in the green dashed box.}}\label{fig: ExpArr_CFD}
\end{figure}

The computational fluid dynamic (CFD) simulations were carried out using a commercial software which is able to solve implicit unsteady Reynolds Averaged Navier-Stokes (RANS) and Large Eddy Simulation (LES) calculations. 
In the case of the RANS simulations, a k-$\omega$ SST two equation turbulence model was used and for the LES a Wall-Adapting Local-Eddy Viscosity (WALE) subgrid scale model was applied to model the turbulent viscosity. 
The computational domain was decomposed into near surface prism layers and a global Cartesian grid with two refinement zones, this can be seen in Fig.\ref{fig: ExpArr_CFD}. 
The initial mesh was based on an anisotropic hexahedral trimmed grid and has 57 million cells for the M-shape and 63 million for the Cupped-shape. 
In order to resolve the fine scale structures within the boundary layer a minimum grid spacing of $y^+<<1$ is used \citep{meyers2008quality}. 
A second-order accurate scheme was selected for the temporal and spatial discretization.
The initial RANS simulations were regarded as converged when the momentum residuals and the energy dropped below $10^{-5}$. 
This criterion was reached within 15,000 iterations.
The LES corresponded to a physical time of 1 second where the time step was $\Delta t = 10^{-5}$.
More information on the stability of the grid and time steps can be found in \citep{Lagemann2018}.
The models used in the simulation were the exact same as the models that were 3D printed, therefore the simulations and experiments could be compared. 

In addition to supporting the experimentally measured lift force the main benefit from the high fidelity simulations was that it allowed for the calculation of the contribution of a particular section or part of the bird to the overall aerodynamic forces. 
The forces were categorized into 3 categories; wing forces (which were split into left and right wing), body forces, and tail forces.
This could be achieved by integrating the static pressure and the tangential stress over the corresponding  area and that would give the resulting normal, axial and side force, $F_y$, $F_x$ and $F_z$ respectively. 
These sectional aerodynamic forces are shown on Fig.\ref{fig:Forces_section} (non-dimensionalized by dividing by the corresponding sectional planform area)

\noindent\subsection{Wind tunnel test.}
The wind tunnel experiments were conducted with a freestream velocity, $U_\infty = 22.5~ms^{-1}$, which was the speed attained by the falcon while flying at equilibrium during the field-experiment at the Oleftal dam in Hellenthal, Germany \citep{ponitz2014}. 
Force measurements were carried out in a G\"{o}ttingen-type wind tunnel at TU Bergakademie Freiberg in 2013 \citep{Colditz:Thesis:2013}, which has a nozzle outlet cross-section of 0.30~m$^2$ (0.6~ x 0.5~m) and a turbulence intensity of 0.04\%.
The falcon models were placed in the middle of the cross-section of the tunnel and mounted to a bespoke force balance via a sting. The force balance consists of three HBM PW15AH load cells which have an accuracy of 0.02\%. 
The forces from the load cells can then be resolved such that the lift and drag can be obtained. More details on the experimental set-up and the calibration of the system can be found in ref. \citep{ponitz2014}. 
In order to compare with well-established relations the forces had to be expressed in non-dimensional form, $C_L$ and $C_D$, by dividing by the corresponding sectional planform area, $S$,  the freestream dynamic pressure, $q_\infty$.

\begin{equation}
C_{(x,y,z)}=\frac{F_{(x,y,z)}}{q_\infty\;S}
\label{eqn: ForceCoeff}
\end{equation}
Considering the forces acting over the horizontal {and normal} plane the lift coefficient, $C_L$ and drag coefficient $C_D$ {respectively} can be expressed as functions of axial and normal force coefficients $C_x$ and $C_y$ and the angle of incidence $\alpha$

\begin{equation}
C_L = C_y\;cos\alpha-C_x\; sin \alpha
\label{eqn: LiftCoeff1}
\end{equation}
\begin{equation}
C_D = C_y\;sin\alpha+C_x\; cos \alpha
\label{eqn: DragCoeff}
\end{equation}

Further insight into the flow topology was obtained with the use of surface oil flow visualisation techniques.
The oil flow visualizations were carried out in a closed loop low speed wind tunnel at City, University of London. The mixture for the surface oil-flow visualisations consisted of Day-Glo powder, white spirit (Naphtha) and Oleic acid.
This was then applied to the model, using a paint brush, immediately prior to the wind tunnel being turned on.
When the wind tunnel is turned on the oil is evaporated by the air flow over the model, leaving the pigment behind on the surface of the model.
The pigment that is left behind shows a `streaky' pattern which shows the near wall streamlines, caused by the shear stress between the surface and the near wall flow. 
Images were subsequently taken under ultra-violet lights in order to improve the contrast of the streaks.





\pagebreak
\section{Results and Discussion}

\subsection{Static aerodynamic loads}

\begin{figure}[b]
\begin{subfigure}[t]{.32\textwidth}
  \centering
  \includegraphics[width=1\linewidth]{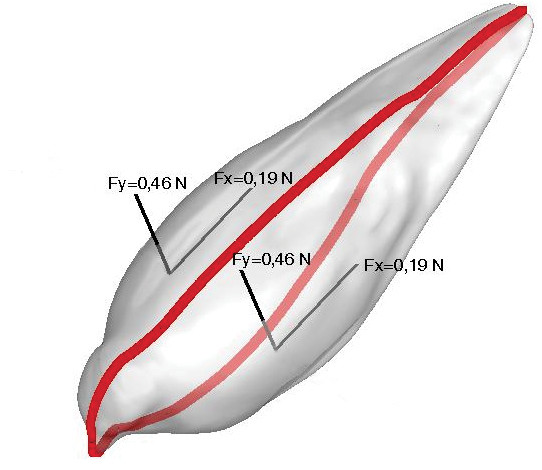} 
\caption{Teardrop-shape}\label{fig: Sectional_T}
\end{subfigure}
\begin{subfigure}[t]{.32\textwidth}
  \centering
  \includegraphics[width=1\linewidth]{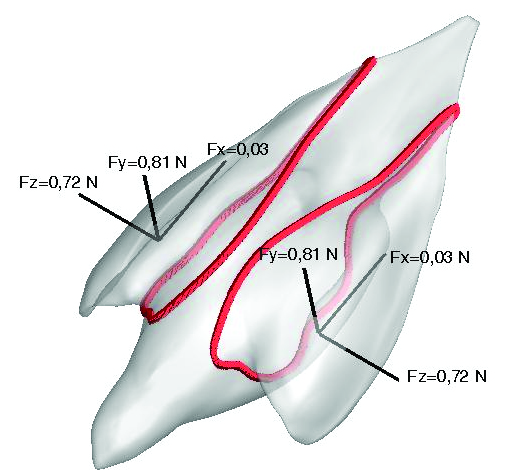}  
  \caption{Cupped-shape}
  \label{fig: Sectional_C}
\end{subfigure}
\begin{subfigure}[t]{0.32\textwidth}
  \centering
  \includegraphics[width =1\linewidth]{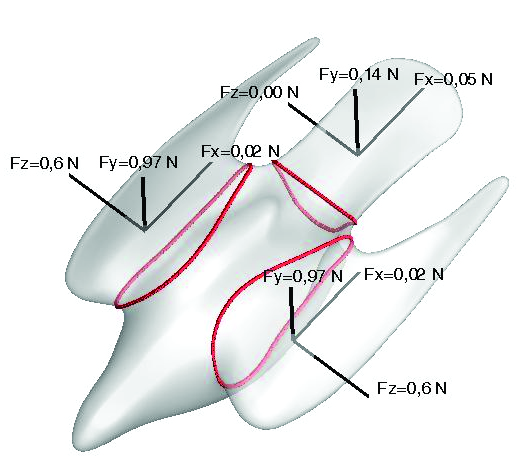}
  \caption{M-shape}
  \label{fig: Sectional-M}
\end{subfigure}
\caption{\textbf{The sectional aerodynamic forces over the falcon, the drag lift and side force are represented as $F_x$, $F_y$ and$F_z$ respectively. All forces shown were taken at $\alpha=5^\circ$.}}
\label{fig:Forces_section}
\end{figure}

The falcon generates most of the lift while in the M-shape, Fig. \ref{fig: Sectional-M}, which is the configuration adopted towards the end of the stoop as shown schematically in Fig. \ref{fig:Flight_path}.
This would help it to start pulling-out by pitching up, immediately before or after striking the prey.
From the field experiment by \citet{ponitz2014}, while flying in the Teardrop-shape and Cupped-shape, during the dive the falcon is normally flying at an angle of incidence, $\alpha<5^\circ$ and therefore generates very little lift as observed from Fig \ref{fig: Sectional_T}.
In these configurations the requirement for lift is minimal.
The bird only wants to achieve maximum speed while in the Teardrop-shape and in the Cupped-shape it is either correcting its attitude to follow a moving prey, by changing the yaw or roll angle or by reducing its speed.
This trend in lift force was also confirmed from the numerical simulation which showed an excellent agreement with the lift measured experimentally on the M-shape.

\begin{figure}[h!]
  \centering
  \includegraphics[width=0.46\linewidth]{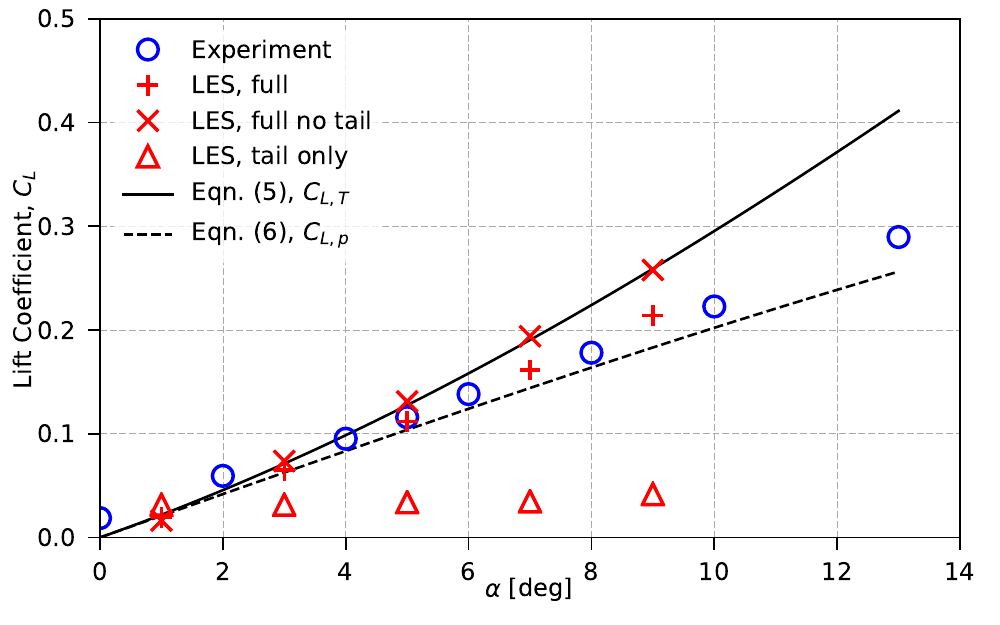} 
 \caption{\textbf{}}\label{fig:lift}
\caption{\textbf{Theoretical, experimental and computational lift coefficient curves; a confirmation that the lift coefficient generated by the wings only in the M-shape agrees well with the lift theory on delta wings.}}
\end{figure}

\begin{figure}[htb!]
 \centering 
\includegraphics[width=0.6\linewidth]{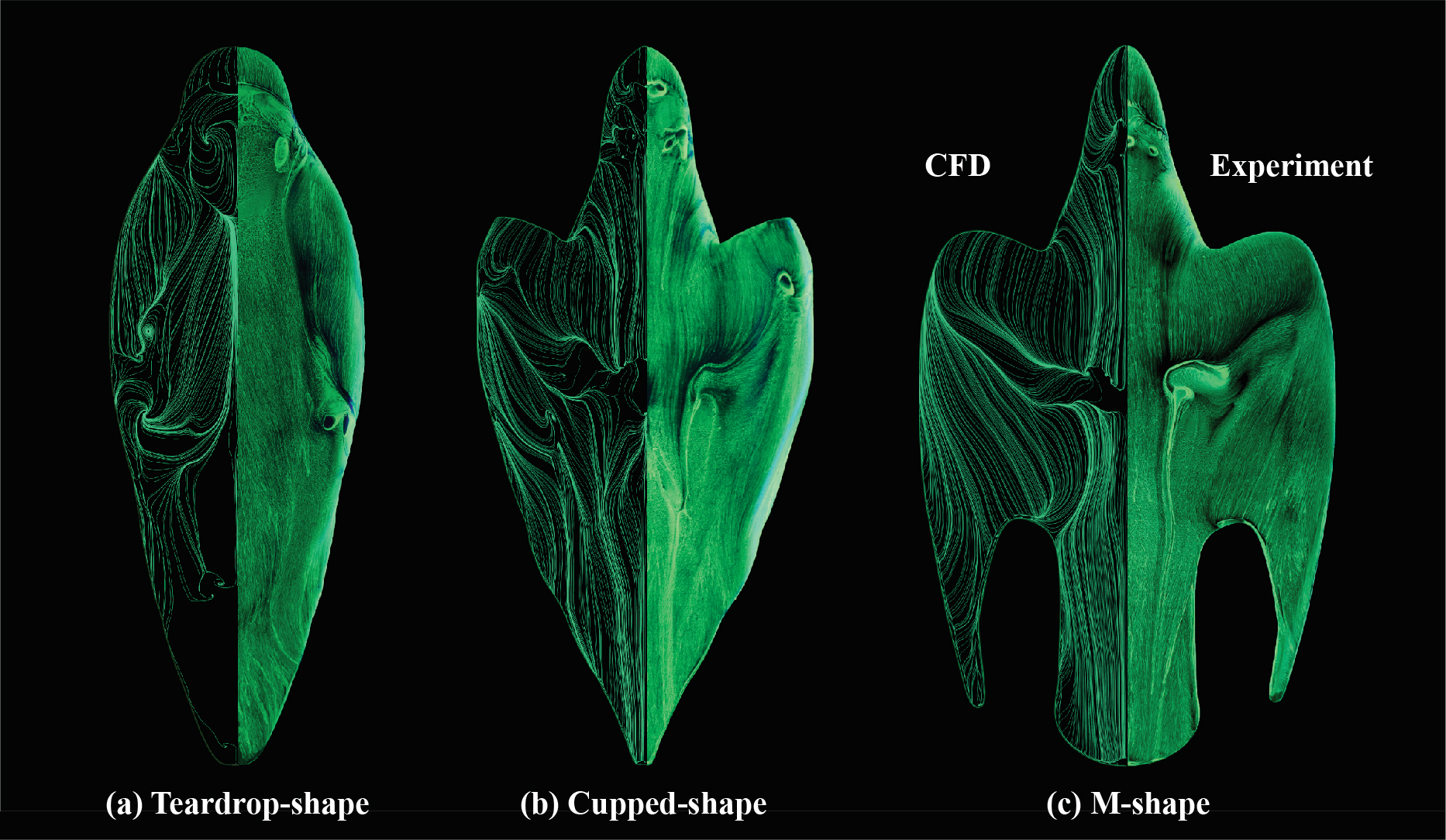}
 \caption{\textbf{Surface streamline patterns from the oil flow visualisation experiments (right half) and the skin friction streaklines from the LES simulation (left half). All images shown taken at $\alpha=5^\circ$.}}
\label{fig:Flow_viz}
\end{figure}

The agreement between the surface oil flow visualisation and the near-wall streamline pattern, at $\alpha = 5^\circ$, seen in Fig. \ref{fig:Flow_viz} confirms the reliability of the simulations.
As reported previously by \citet{Gowree2018} and shown by the highly 3D and inflectional surface shear stress lines in Fig. \ref{fig:Flow_viz} the flow over the falcon is dominated by vortical structures, even in the low lift, Teardrop-shape and Cupped-shape configurations. The linear characteristic of the total lift from the M-shape in Fig. \ref{fig:lift} encouraged further comparison with well established lift theory.

From the theory developed by \citet{Polhamus1971}, in a subsonic flow the total theoretical lift, $C_{L_T}$, generated on a delta wing can be given as the sum of the potential flow lift $C_{L_p}$ and the contribution from the vortex flow, $C_{L_v}$. Therefore, 
\begin{eqnarray}
C_{L_T}  &=& C_{L_p} + C_{L_v} \label{eqn: CL}\\
C_{L_p} &=& K_p\, sin \alpha \, cos^2 \alpha \label{eqn: CLp}\\
C_{L_v} &=& K_v\, sin^2 \alpha\, cos \alpha 
\end{eqnarray} 

\noindent where $K_p$ and $K_v$ are the potential flow and vortex flow  coefficients, given as a function of wing aspect ratio. For the comparison, these values were taken as $K_p=1.2$ and $K_v=\pi$, where the aspect ratio of an equivalent delta wing was taken as $AR=1$.
\pagebreak

Since the flow over the bird was dominated by large vortices similar to that on delta wings, Polhamus' \citep{Polhamus1971} theory for unsteady lift generation was preferred as opposed to the steady thin airfoil theory. In Fig.\ref{fig:lift} at first glance, at low incidence, $\alpha$, both the lift coefficient, $C_{L_T}$, measured and computed on the full model showed a closer agreement with the potential flow theory, $C_{L,p}$. Further increase in $\alpha$ showed a deviation, but still not close enough to the theory for vortex lift. After subtracting the lift generated by the tail from the overall lift and following the appropriate non-dimensionalization, the agreement with the vortex lift theory in Fig. \ref{fig:lift} is significantly improved.

Despite the formation of the strong vortices on the tail, the lift coefficient does not match the theory of vortex lift in Fig. \ref{fig:lift} and therefore the tail does not act as a delta wing, an observation supported by previous investigations \citep{Maybury2001,Evans2002,Evans2003}. In fact, the tail generates lift even at low angle of incidence, at $\alpha \approx 0^\circ$.
This ensures that it can be used for control during any flight conditions.
From Fig. \ref{fig:Flow_viz}, especially on the M-shape, it is important to point out that there is an upstream conditioning of the flow in the mid-dorsal region.
This promotes a reattachment to ensure that the flow does not separate while progressing towards the tail and hence high effectiveness of the tail is maintained and major deployment is not necessary.
This has been reported in detail by \citet{Gowree2018}.

\subsection{Static longitudinal stability}
\begin{figure}[h]
\begin{subfigure}[t]{.495\textwidth}
  \centering
  \includegraphics[width=1\linewidth]{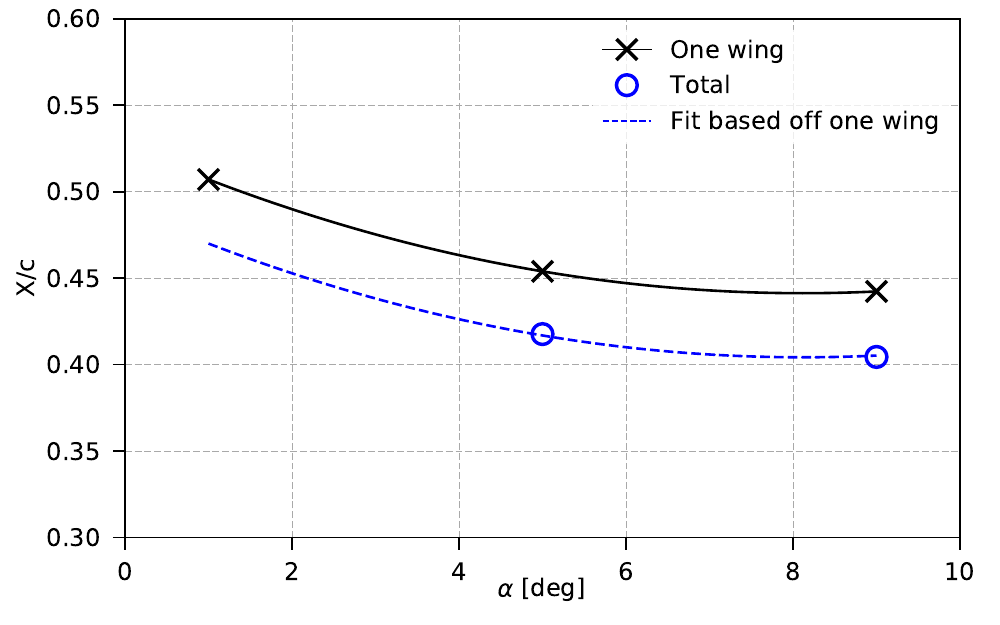} 
\caption{\textbf{The CoP location along the bird's longitudinal axis}}
\label{fig:CoP}
\end{subfigure}
\begin{subfigure}[t]{.495\textwidth}
  \centering
  \includegraphics[width=1\linewidth]{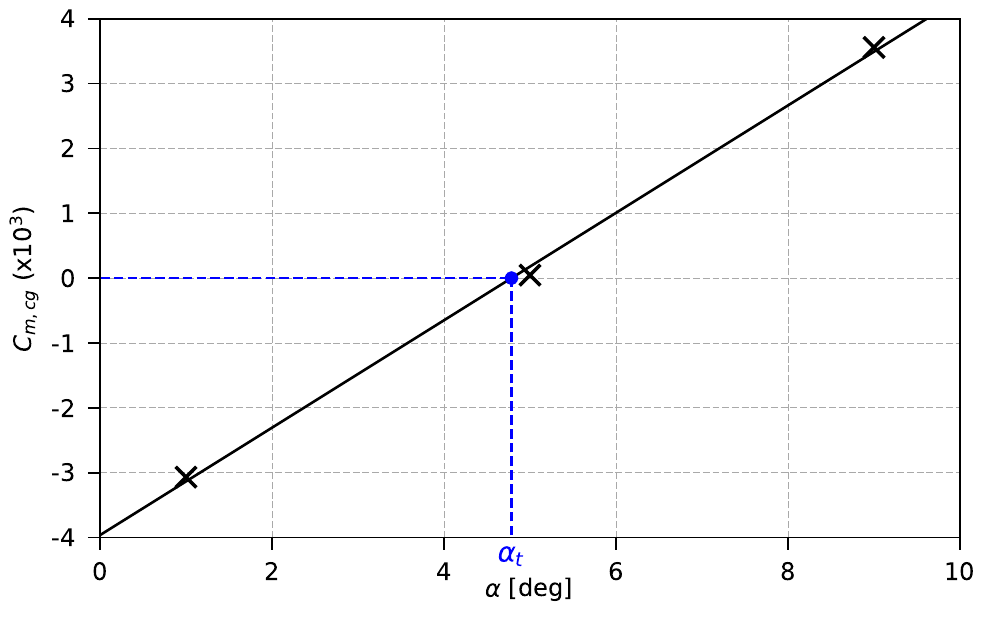}  
  \caption{\textbf{Pitching moment coefficient about the CG for the wing-body-tail combination, where $\alpha_t$ indicates the trim angle}}
  \label{fig: CM_Alpha}
\end{subfigure}
\caption{\textbf{CoP location and resulting pitching moment analysis}}
\label{fig:CM_Alpha&CoP}
\end{figure}

\begin{figure}[h]
\begin{subfigure}{.495\textwidth}
  \centering
  \includegraphics[width=1\linewidth]{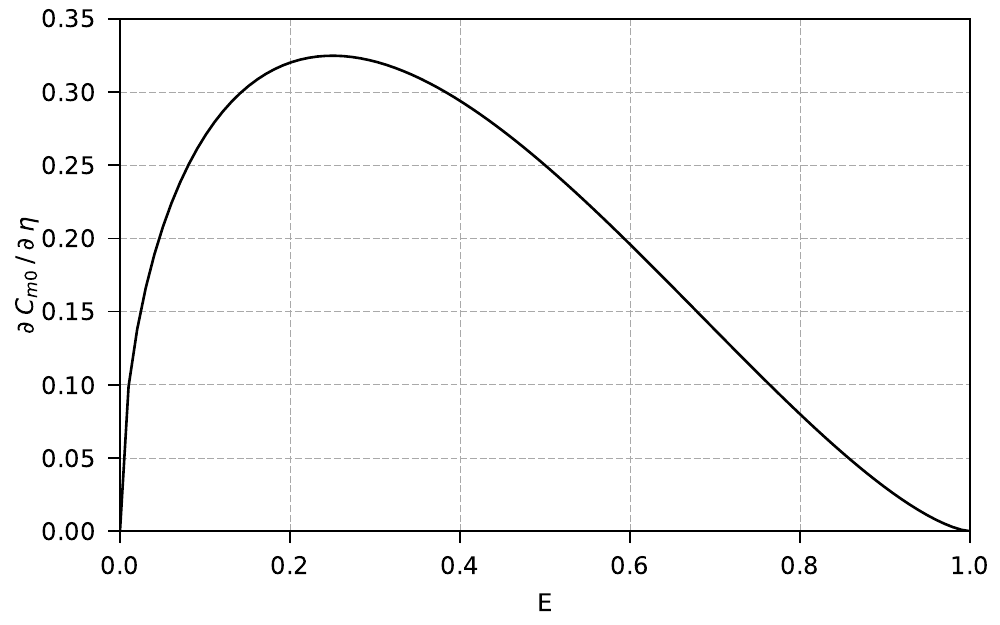} 
\caption{\textbf{}}
\label{fig: Flap-to-chord}
\end{subfigure}
\begin{subfigure}{.495\textwidth}
  \centering
  \includegraphics[width=0.65\linewidth, angle = -90]{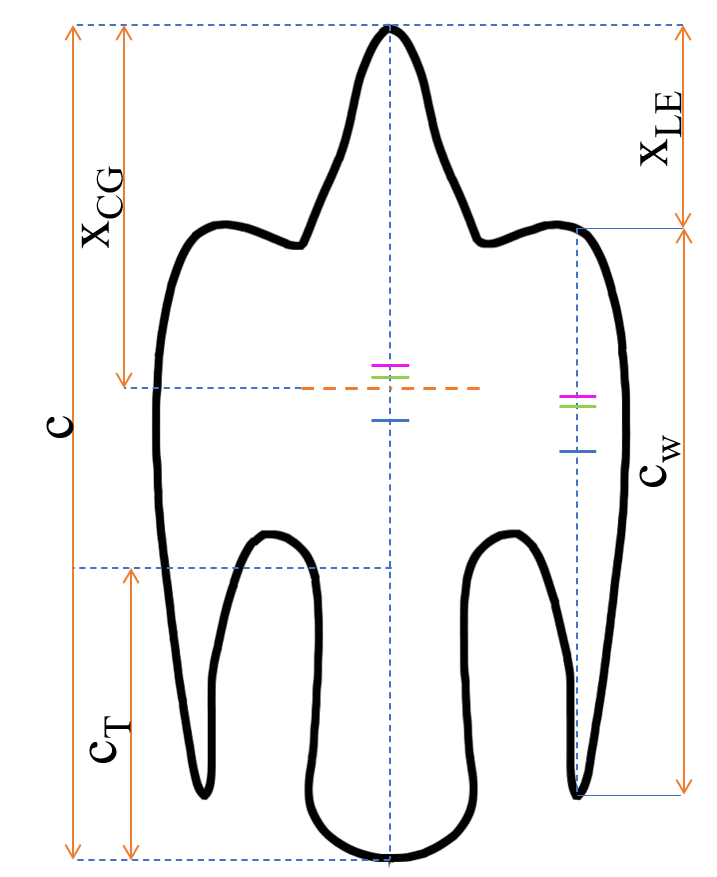} 
  \caption{\textbf{}}
  \label{fig:dims}
\end{subfigure}
\caption{\textbf{(a) Change in pitching moment coefficient $C_{M0}$ per angle of flap deflection $\eta$ as given by \citet{glauert1927theoretical}. (b) Outline of model showing lengths and CoP locations; where (\protect\begin{tikzpicture}
\protect\tikz[baseline=-1pt] \protect\draw[solid,very thick,color=cyan] (0.0,0.0) -- (0.35,0.0);
\protect\end{tikzpicture}) indicates the location at $\alpha=1^\circ$, (\protect\begin{tikzpicture}
\protect\tikz[baseline=-1pt] \protect\draw[solid,very thick,color=lime] (0.0,0.0) -- (0.35,0.0);
\protect\end{tikzpicture}) at $\alpha=5^\circ$ and (\protect\begin{tikzpicture}
\protect\tikz[baseline=-1pt] \protect\draw[solid,very thick,color=magenta] (0.0,0.0) -- (0.35,0.0);
\protect\end{tikzpicture}) at $\alpha=9^\circ$. Abbreviations and subscripts outlined in Nomenclature section}}
\end{figure}

For comparison, the center of pressure (CoP) was first built from only the forces acting on one wing and compared to that built from the entire wing/body/tail combination, Fig. \ref{fig:dims} and \ref{fig:CoP}.
This allows conclusions to be drawn about the importance of the body/tail on the stability. 
The positively cambered wings will generate a negative `nose-down' pitching moment by their nature.
The same is true for the body and tail of the bird, as they are also lift producing surfaces, contrary to tail-planes in most aircraft.
The pitching moment about the aerodynamic center was found by tracking the movement of the CoP at various incidences and taking the moment about various points until a location, $x_{ac}$, was found where the pitching moment stayed fairly constant. For the pitching moment around the the center of gravity (CG) it follows:

\begin{equation}
 C_{m,cg} = C_{m_0,ac}+C_y\cdot \left( \frac{x_{cg}-x_{ac}}{c_w}\right)
 \label{eqn: MomentCoefCG1}
\end{equation}
where $C_{m_0,ac}$ is the zero lift pitching moment of the body about its aerodynamic center and $C_y$ is the normal force coefficient of the body.

Classical longitudinal stability theory states that the criteria for stability are that there exists an appropriate angle of attack at which the aircraft is in pitching moment equilibrium (trimmed flight) and the pitching moment of an aircraft after departure from said equilibrium always be restoring.
That is, that the moment be negative (pitch down) following an upwards disturbance and positive (pitch up) following a downwards one.
This can be quantified by inspecting the slope of the pitching moment curve at trimmed condition (ie. $C_{m,cg}=0$).
The condition is that the slope be negative. That is,
at an incidence $\alpha$ = $\alpha_t$,  $C_{m,cg}=0$, and,
\begin{equation}
    \frac{d C_{m,cg}}{d \alpha} < 0
\end{equation}

Analysing the forces and moments from the LES results can give an insight into the longitudinal stability of the falcon in such a maneuver.
The center of pressure for one wing shows, as expected, a steady movement towards the leading edge with the increase in incidence.
Similarly, this was seen in that of the entire bird, however, the point at 1 degree was shown to be an outlying point and subsequently, using the fit based on the wing CoP movement, the point at 1 degree was estimated as shown in Fig. \ref{fig:CoP}

Using equation \ref{eqn: MomentCoefCG1}, the pitching moment about the CG of the model was plot (Fig. \ref{fig: CM_Alpha}) and shown to have a marginal positive (unstable) slope.
The trim point was found at ~ $\alpha_t \approx 5^\circ$.
This is close to the angle of attack observed in Peregrine falcons stooping flight at the start of the pull-out \citep{ponitz2014}.
This is the angle of attack at which, at the tail setting of the model, the bird would aim to fly in order to maintain balanced equilibrium of moments about the CG.
The trim point is usually adjusted by tail morphing and this reflects that of the current (nominal) tail position.

The limited use of the tail for such adjustments at high speed could also be attributed to its high effectiveness, such that small deflections can produce large pitching moments without the need to fully deploy, this is due to the ratio of the tail-to-body length, acting as a trailing-edge flap.
For a plain-flap-airfoil, the change in pitching moment coefficient $C_{M0}$ per angle of flap deflection $\eta$ is illustrated in Fig. \ref{fig: Flap-to-chord} and given by \citet{glauert1927theoretical} as:
\begin{equation}
    \frac{\partial C_{m0}}{\partial \eta}=\sqrt{E(1-E)^3}
\end{equation}
where E is the flap-to-chord ratio for an airfoil.
Considering the tail as a trailing-edge flap, the tail length of the falcon model $C_T$ is $\approx$ just over a third of its body length $C$, Fig. \ref{fig:dims} as per the sectional forces taken in Fig. \ref{fig: Sectional-M}.
This would correspond to a value of $E=0.35$ in Fig. \ref{fig: Flap-to-chord} where it is most effective (near the maximum at the curve).
This natural sizing of optimized trailing edge morphing surfaces was also found in aquatic life in hydrodynamic analysis of trout fish \citep{przybilla2010entraining, brucker2006dynamic}.
Treating the tail as a 'trailing-edge' control surface, such as an elevon or flaperon, it was found that Barn Owls use downward tail deflection for lift production to offset the loss of continuous spanwise lift over the body and therefore reduce induced drag \cite{Song2020Virtual}. The tail-length to body-length ratio of Barn Owls is $\approx 0.38$ for adults \cite{marti1990sex}, slightly larger than that of the Peregrines but still within the highly effective area.  This suggests they also seem to have naturally optimized tail sizes for pitching control.

It is apparent that, in a pull-out maneuver, the falcon will fly in a marginal longitudinal static instability.
This could be to exploit the benefits in maneuverability in addition to eliminating trim drag - the portion of drag associated with lift produced to counter tail balancing in stable aircraft.

\pagebreak

\subsection{Loads during pull-out}

The pull-out for an ideal falcon is defined by \citet{Tucker1998b} as the phase in which it follows a circular arc until the glide path is horizontal.
In reality, this path is a series of arcs decreasing in radius as the bird nears horizontal.
By taking a centrifugal force balance about the center of the instantaneous circular pull-out arc, it can be shown that, in order to sustain the maneuver, the lift is equal to the sum of the centrifugal force and the component of weight radial to the pull-out (ie. normal to the flight direction).

Using the instantaneous pitching angle ($\theta$) and velocity ($U$) from the recordings \citep{ponitz2014}, for the weight of the bird in question, the load factor ($n$) can be obtained via equation \ref{eqn: LFactor} and subsequently the aerodynamic loads.
\begin{equation}
n = \frac{U\dot{\theta}}{g} + cos(\theta)
\label{eqn: LFactor}
\end{equation}

\begin{figure}[h!]
\begin{subfigure}[t]{.5\textwidth}
  \centering
  \includegraphics[width=1\linewidth]{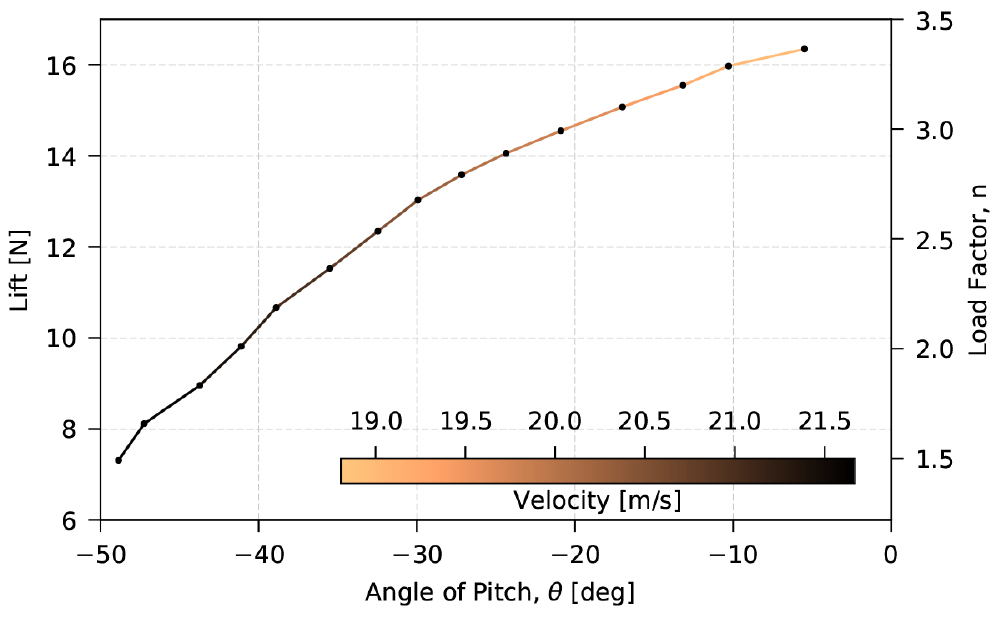} 
 \caption{\textbf{}}
 \label{fig:trajectorylift}
\end{subfigure}
\begin{subfigure}[t]{.41\textwidth}
   \centering
    \includegraphics[width=1\linewidth]{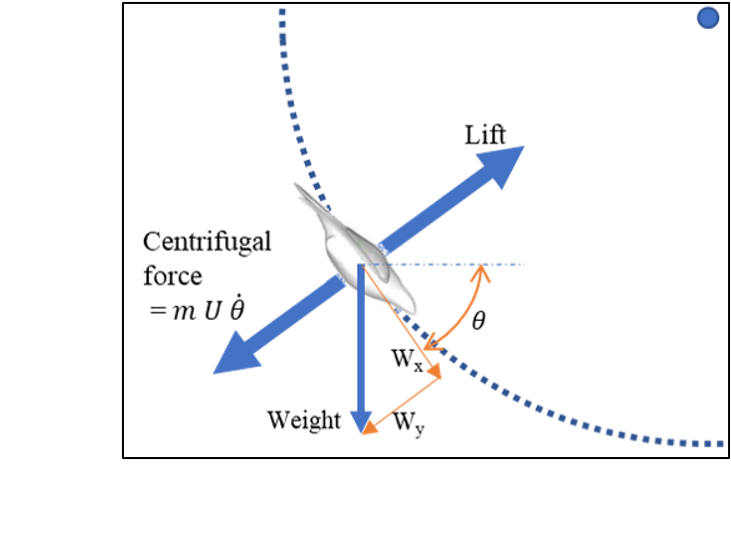}
    \caption{\textbf{}}
    \label{fig:forcebalance}
\end{subfigure}
\caption{\textbf{(a) Lift and load factor throughout a pull-out maneuver. 
(b) Sketch illustrating centrifugal force balance during a pull-out.
These were obtained by using equation \ref{eqn: LFactor} considering the values reported from the field-experiment by \citet{ponitz2014}}}\label{fig:TrajectAnalysis}
\end{figure}

In Fig. \ref{fig:TrajectAnalysis} a gradual increase in lift is observed throughout the maneuver as the bird tightens its pull-out radius and approaches level flight. While the velocity can be seen to undergo a marginal decrease to prevent the bird from stalling or allowing it to go around for another strike in case if it misses the prey. In this case a maximum load factor of 3.2g was estimated despite the field experiment being conducted on a trained falcon in a controlled environment, where the maximum speed attained was limited by the altitude  of the dam in ref. \citep{ponitz2014}

The velocity attained during the field experiment in ref \citep{ponitz2014} was significantly lower than that normally observed on wild falcons in stoop, which seem to be the highest during courtship stoop demonstration \citep{Bleckmann2015}. Peregrine falcons can reach velocities of $80 ms^{-1}$ and by setting this as the initial velocity for the pull-out (ie. the velocity at equilibrium), using the same methods as before, the bird will produce up to a maximum of $47 N$ of lift at the tightest part of the pull-out.

\subsection{Wing morphing during pull-out}
\begin{figure}[h!]

\begin{subfigure}[t]{.495\textwidth}
  \centering
\includegraphics[width=0.9\linewidth]{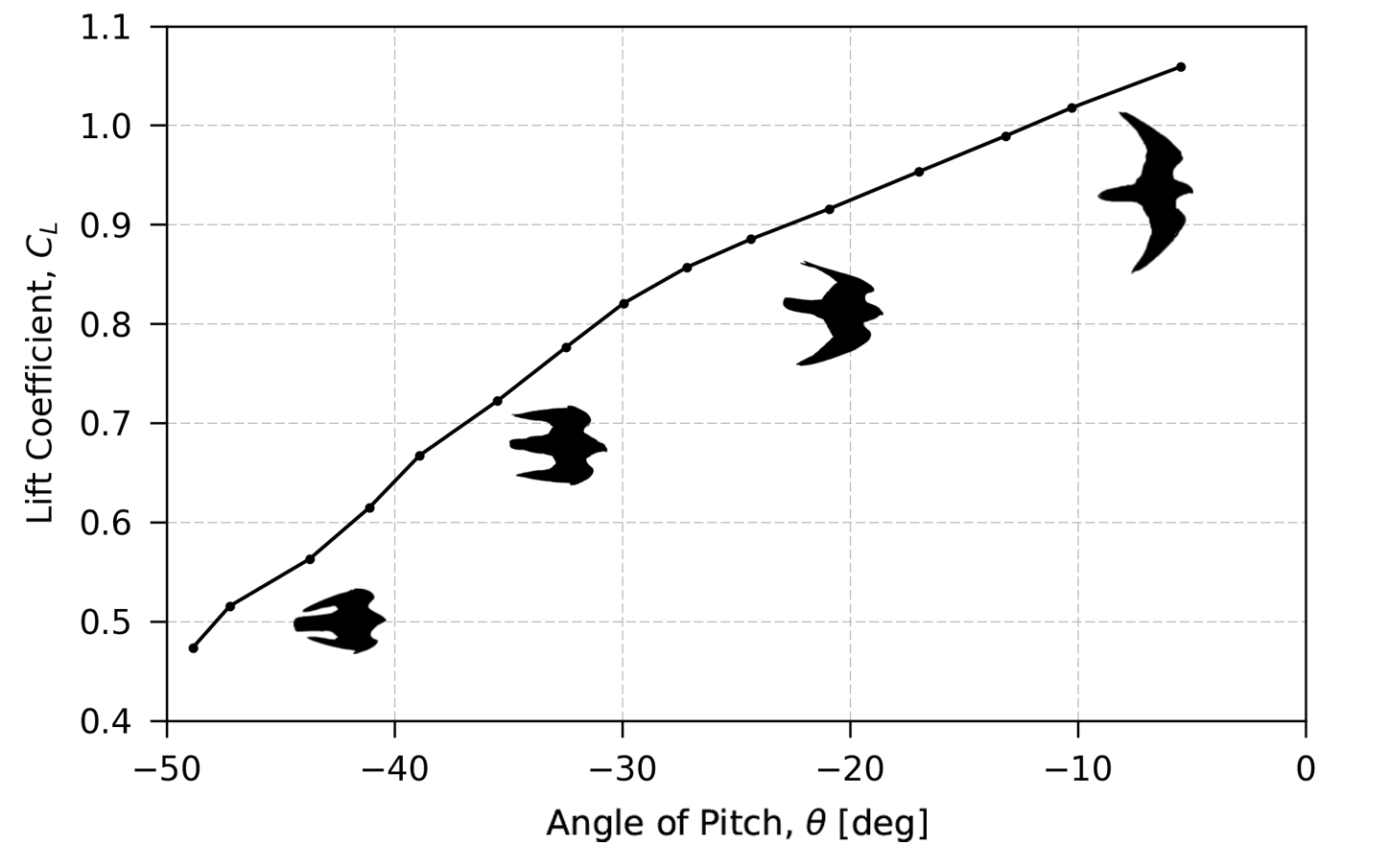}
 \caption{\textbf{}}
\label{fig:trajectoryliftcoef}
\end{subfigure}
\begin{subfigure}[t]{.495\textwidth}
  \centering
  \includegraphics[width=0.9\linewidth]{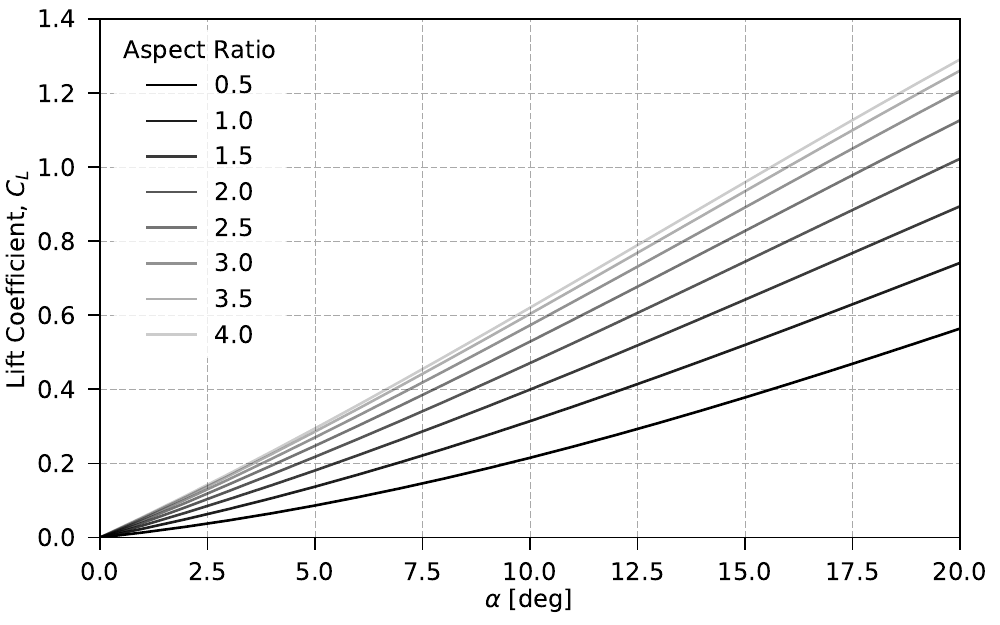} 
 \caption{\textbf{}}
 \label{fig:Aspect Ratio}
\end{subfigure}
\caption{\textbf{(a) Lift coefficient in a pull-out illustrating the different configurations. (b) Illustration of the effect of varying the aspect ratio $(AR)$ on the theoretical lift coefficient.}}
\end{figure}

{In the previous section the lift generated by the bird during the pull-out was calculated based on the glide path and speed of the bird during the field experiment. After non-dimensionalization, from Fig. \ref{fig:trajectoryliftcoef}, the maximum $C_L$ reached in the M-shape was twice higher than that measured during the wind tunnel test or computed from the LES simulations, even at high incidence. In order to attain such high $C_L$ while maintaining the M-Shape the bird had to be flying at very high incidence, which would increase the form drag significantly, but Fig. \ref{fig:TrajectAnalysis} confirms that the reduction in speed in this phase was marginal. To attain such high lift coefficient without significant increase in drag the bird was seen to modify its wing morphology during the pull-out, following the trends represented by the sequence of silhouettes in Fig. \ref{fig:trajectoryliftcoef}, which was confirmed from the live recordings during the field experiments. Here, the main morphological transformation occurs through the sweep angle change of the primaries which would result in an increase in span and lifting surface. While shifting from the maximum sweep angle, $\Lambda_{max}=90^\circ$, from the M-shape to $\Lambda_{min}=40^\circ$, in the deployed wings shape towards the end of pull-out, the aspect ratio, $AR$, for the model is increased from approximately 0.95 to 4.9 respectively. Earlier it was demonstrated that the lift generation by the wings was similar to that over delta wings. Extension of the delta wing theory here for a range of aspect ratios, Fig. \ref{fig:Aspect Ratio} confirms that in order to achieve maximum required $C_L$ towards the end of pull-out while operating at moderate to low angle of incidence an aspect ratio tending towards 5 at least will be required. From Fig. \ref{fig:Aspect Ratio} it can be seen that as the aspect ratio approaches $AR = 4$, the curves seem to converge.
Hence it can be concluded that when $AR>4$, the $C_L$ curves will collapse on to one another.}

{Although the increase in lift is achieved through in plane symmetrical morphing (change in sweep) of primary feathers moving from the M-Shape towards fully deployed primaries, both in-plane and out-of-plane asymmetrical morphing provides yaw control coupled with roll, through a bank turn in the absence of a vertical tail-plane unlike in most aircraft. This asymmetrical morphing is also observed during other phases of the stoop when the bird is re-adjusting it's trajectory. Pure yaw control in the Cupped-shape can be achieved due to the substantial amount of side force generated on the wings as shown in Fig. \ref{fig:Forces_section} (compare \citep{ponitz2014a}), also by the strong vortices which are now aligned to the side of the bird. This side force does not degrade significantly while morphing into the M-shape and hence allowing the bird to engage easily into a yaw maneuver, if it needs to turn around for another attempt.}

Lacking vertical stabilizers, birds need to continuously augment their yaw stability by twisting their tails \cite{thomas2001animal}, faster birds need larger tails for this purpose as seen in swifts and hawks \cite{brown1963flight}. This element of side-force generated by the out-of-plane morphing can offer an element of yaw control in a crabbed turn in addition to possibly augmenting yaw stability via asymmetric morphing reducing the load, and the associated size constraint, from the tail in a pull-out or eliminating its need entirely. These vortices are the main contributors to the aerodynamic forces and play an important role in the manoeuvrability of the falcon, where the pitch, yaw or roll could be achieved simply by morphing and projecting the wing over different planes.

\pagebreak


\section*{Conclusion}
{An assessment of the static aerodynamic loads generated by the bird in the M-shape showed that the lift generated purely by the wing matches very well with Polhamus' theory and this is a result of the formation of large vortices, similar to that on delta wings.
The extension of the delta-wing theory allowed for further flight mechanics analysis of the bird towards the end of the stoop or pull-out.
The static longitudinal stability analysis confirmed that, in the M-shape configuration, the bird was flying unstably in pitch, based on the positive slope in the pitching moment coefficient.
This allows the falcon to fly in a responsive and maneuverable fashion and in a trajectory with minimal loss in energy and forward velocity.
Further analysis of the pull-out maneuver showed that during the controlled field experiment of \citet{ponitz2014} the bird would be experiencing a load factor of approximately 3.5g, but can be tripled during stoops in wild conditions.}

{Analysis of the wing morphing showed that the high lift required towards the end of pull-out was achieved by forward sweeping of the primary feathers while maintaining the angle of incidence as low as possible for lower drag.
Asymmetrical and out of plane morphing of the primaries in the form of dihedral change will also provide yaw and roll control without much input from the tail.
The furling of the tail in M-shape suggests that this control is handled via instantaneous morphing of the wings - ie. changing the sweep/aspect ratio to alter the amount of lift and pitching moment to match the requirement, reducing or eliminating the need for large tail inputs and their associated increase in drag. The minimal requirement of tail inputs for longitudinal adjustments was also attributed to its high effectiveness, this is due to the natural optimisation of tail-to-body length, observed also in other avian and aquatic species.}

{Some quantitative evidences have been provided herein on how a falcon can achieve its superior maneuverability through wing-morphing, from live-flight observations and measurements during field experiments, wind tunnel testing and numerical simulation. These findings can be extrapolated for the design of convertible bio-inspired UAVs and MAVs or even more versatile and highly maneuverable fixed-wing aircraft.}  

Incorporating this in bio-inspired UAV/MAV's could be realized with vortex-lift/morphing wing aircraft controlled by a light-weight FCC and suite of sensors designed to detect departures from equilibrium and respond actively and instantaneously with simple wing-morphing to maintain desired attitude in a pull-out maneuver or any similar maneuver.
In order to design this, more needs to be known about the effect of altering sweep and aspect ratio on leading-edge vortex lift production and center of pressure location for small-scale, high-speed, morphing-wing aircraft and natural fliers at Reynolds numbers $\approx 10^5$ to $10^6$.
The benefits of adopting nature-inspired control alternatives are a possible reduction in the number of control surfaces, their weight, response time, and their associated increase in power consumption and overall drag.

\pagebreak

\section*{Funding Sources}
The position of Professor Christoph Br\"{u}cker is co-funded by BAE SYSTEMS and the Royal Academy of Engineering (Research Chair no. RCSRF1617$\backslash$4$\backslash$11), and the position of Omar Selim is funded by the George Daniels Educational Trust, all of which are gratefully acknowledged.
The flow visualization experiments carried out at City, University of London were part funded by the National Wind Tunnel Facilities grant (NWTF grant number: EP/L024888/1).

\section*{Acknowledgments}
The authors express profound thanks to T. Colditz and M. Gelfert by whom the force measurements were carried out on the different falcon models in 2013 as part of their studies at TU Bergakademie Freiberg. In addition, we thank Mike Newsam from Stellar Advanced Concepts Ltd for his fruitful discussions at City, University of London on bio-inspired wing-morphing concepts. 

Authors CRediT statement; \textbf{O.S}: Conceptualization, Methodology, Validation, Formal analysis, Investigation, Data Curation, Writing - Original Draft, Writing - Review \& Editing, Visualization, Project administration.
\textbf{E.R.G}: Conceptualization, Validation, Formal analysis, Investigation, Writing - Original Draft, Writing - Review \& Editing.
\textbf{C.L}: Methodology, Software, Validation, Formal analysis, Investigation, Data Curation, Writing - Original Draft, Writing - Review \& Editing, Visualization.
\textbf{E.T}: Methodology, Validation, Investigation, Data Curation, Writing - Original Draft, Writing - Review \& Editing, Visualization.
\textbf{C.J}: Methodology, Validation, Investigation, Writing - Original Draft, Writing - Review \& Editing.
\textbf{C.B}: Conceptualization, Methodology, Writing - Original Draft, Writing - Review \& Editing, Supervision, Funding acquisition. 
\bibliography{references}

\end{document}